\documentclass{aa}
\usepackage{txfonts} 
\usepackage{graphicx}% Include figure files
\usepackage{dcolumn}% Align table columns on decimal point
\usepackage{bm}% bold math
\usepackage{latexsym}
\usepackage{epsf}
\usepackage{epsfig}
\def\eqref#1{(\ref{#1})}
\newcommand{\fr}{{^F\hspace{-.02in}R}}
\def\diag{\mathop{\rm diag}}
\begin{document}
\title{Tidal Acceleration of  Ultrarelativistic Particles}
\author{C. Chicone\inst{1}\and B. Mashhoon\inst{2}}
\institute{Department of Mathematics, University of Missouri-Columbia, 
Columbia, Missouri 65211, USA \and Department of Physics and
Astronomy, University of Missouri-Columbia, Columbia, Missouri 65211, USA\\
e-mail: MashhoonB@missouri.edu}
\date{Received/Accepted}
\abstract{
We investigate the motion of free relativistic particles relative to the ambient
medium around a gravitationally collapsed system. 
If the relative speed exceeds a critical value
given by $c/\sqrt{2}$, the gravitational tidal effects exhibit novel features
that are contrary to Newtonian expectations. In particular, ultrarelativistic jet
clumps moving freely outward along the rotation axis strongly decelerate with
respect to the ambient medium, while ultrarelativistic particles strongly
accelerate in directions normal to the jet axis. The implications of these direct
consequences of general relativity for jets in microquasars and the origin of the
high-energy cosmic rays are briefly mentioned.
}
\maketitle
\section{Introduction}
\begin{figure}
\vspace{1in}
\centerline{\psfig{file=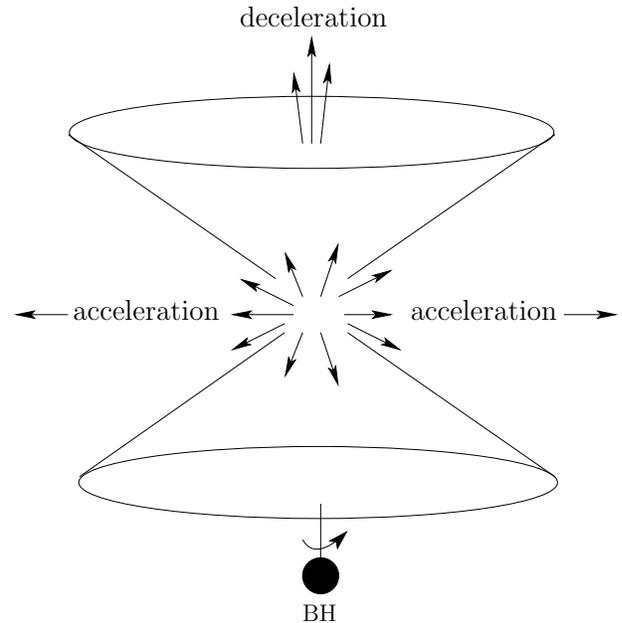, width=12pc}}
\caption{Schematic diagram of the critical velocity cone.  \label{fig:1}}
\end{figure}
The purpose of this Letter is to discuss some of the observable consequences
of the motion of relativistic particles in the field of gravitationally collapsed
configurations. 
In particular, we point out novel features of general relativity 
regarding the tidal
influence of a neutron star or a black hole on the motion of nearby free particles
 moving faster than
the critical speed $c/\sqrt{2}\approx  0.7 c$
 relative to the ambient medium surrounding the
central mass.  Such ultrarelativistic particles originating 
near the poles of the
collapsed system suffer significant tidal  \emph{deceleration}
 within a cone of angle $\theta \approx 55^\circ$
around the axis of rotation. Here $\theta$ represents half the angle
at the vertex of the cone.
Outside the cone, however, an ultrarelativistic
particle experiences tidal \emph{acceleration}.  The situation is depicted
schematically in Fig.~\ref{fig:1}
for the north pole of a Kerr  black hole.  
Similar outflows are expected near the south pole.

The solutions of the geodesic equation as well as the Lorentz
force equation in black hole fields in standard Schwarzschild-like systems
of coordinates have been extensively studied and many
useful results have thereby been obtained (see, e.g., Chandrasekhar~\cite{1}). For
the study of certain phenomena, however, it is more useful to study 
\emph{relative} motion, i.e. motion of one particle with respect to another. 
This is in keeping with the spirit of the theory of relativity.
It also corresponds to the physical interpretation of some 
astronomical situations.
Consider, for instance, the \emph{Chandra} X-ray images of the 
Crab Nebula: to reveal the accelerating rings in the equatorial plane, 
different images have to be compared with each other keeping 
certain central features ``fixed" 
(http://chandra.harvard.edu/photo/2002/0052/index.html).
To study relative motion in an invariant setting corresponding to 
actual observations, it is useful to 
establish a Fermi normal coordinate system along a reference 
geodesic $\mathcal{O}$.
The motion of a nearby free particle with respect to $\mathcal{O}$ 
is then given by the geodesic equation in the quasi-inertial
Fermi coordinate system (Synge \cite{2}). 

Let $\tau$ be the proper time and $\lambda^\mu_{\;\;(0)}=dx^\mu/d\tau$
be the four-velocity vector of  $\mathcal{O}$. 
A triad of ideal
orthonormal gyro directions $\lambda^\mu_{\;\;(i)}$, $i=1,2,3$, can be
parallel propagated along $\mathcal{O}$'s worldline such that 
$\lambda^\mu_{\;\;(\alpha)}(\tau)$ is a local orthonormal tetrad frame. 
Based on these local axes, the Fermi coordinates
$(T,\mathbf{X})$  simply provide a geodesic normal coordinate
system along the worldline of $\mathcal{O}$. 
The reference worldline is then the temporal axis $(T=\tau,\mathbf{X}=\mathbf{0})$
and the metric tensor in the Fermi system is given by
\begin{eqnarray}
\label{1} 
\nonumber g_{00}&=&-1-\fr_{0i0j}(T)\,X^iX^j+\cdots,\\
\nonumber g_{0i}&=&-\frac{2}{3}\,\fr_{0jik}(T)\, X^jX^k+\cdots,\\
g_{ij}&=&\delta_{ij}-\frac{1}{3}\,\fr_{ikj\ell}(T)\, X^kX^\ell+\cdots,
\end{eqnarray}
where
\begin{equation}\label{2}
\fr_{\alpha \beta \gamma \delta 
}(T)=R_{\mu \nu \rho \sigma }\lambda^\mu_{\;\; (\alpha )}\lambda^\nu 
_{\;\;(\beta)} \lambda^\rho _{\;\; (\gamma )}\lambda^\sigma 
_{\;\;(\delta )}
\end{equation}
is the projection of the Riemann tensor on the local frame 
of $\mathcal{O}$. 
Along $\mathcal{O}$'s worldline, the metric reduces to the Minkowski
metric according to~\eqref{1}. The deviations from 
an inertial system of coordinates are quadratic and higher order
in distance away from the reference worldline.
The Fermi coordinates are admissible only in a cylindrical region
of radius $ \mathcal{R}(T)$
around the reference worldline, where $\mathcal{R}$
is a certain minimum radius of curvature determined via the Riemann
tensor and its covariant derivatives along the worldline of  $\mathcal{O}$.
The equations of general relativity may be expressed in any admissible
coordinate system. We are interested in the geodesic equation of motion in
the quasi-inertial Fermi coordinate patch.

\section{Generalized Jacobi equation}

The geodesic equation for a free test particle $\mathcal{P}$ in the Fermi system can be 
reduced to the general tidal equation (Mashhoon \cite{3,4}); however,
if we limit our considerations to the terms
given explicitly in equation~\eqref{1},
we get the generalized Jacobi equation (Chicone \& Mashhoon  \cite{5})
\begin{eqnarray}\label{3}
\nonumber &&\frac{d^2X^i}{dT^2}+\fr_{0i0j}X^j+2\,\fr_{ikj0}V^kX^j\\
&& 
+\frac{2}{3}(3\,\fr_{0kj0}V^iV^k+\fr_{ikj\ell}V^kV^\ell 
+\fr_{0kj\ell}V^iV^kV^\ell )X^j=0,
\end{eqnarray}
where $\mathbf{V}=d\mathbf{X}/dT$ is such that the 
four-velocity of $\mathcal{P}$ 
is given by $U^\mu = \Gamma ( 1, {\bf V})$
 and $\Gamma$ is the modified Lorentz factor given by
\begin{eqnarray}\label{eq37}
\nonumber \frac{1}{\Gamma^2}&=& 1-V^2+\fr_{0i0j}X^iX^j 
+ \frac{4}{3} \fr_{0jik}X^jV^i X^k\\
&&{}+ \frac{1}{3} \fr_{ikj\ell} V^iX^kV^jX^\ell
\end{eqnarray}
at the level of approximation under consideration here.
For $|\mathbf{V}|\ll 1$, we may neglect all terms proportional to 
the relative velocity in the equations of motion and \eqref{3} reduces to the
familiar Jacobi equation (Synge \cite{2})
\begin{equation}\label{4}
\frac{d^2 X^i}{dT^2}+K_{ij} X^j=0.
\end{equation}
Here $K_{ij}=\fr_{0i0j}$
is the symmetric tidal matrix that reduces in the Newtonian limit
to $\partial^2\Phi/\partial X^i\partial X^j$,
where $\Phi$ is the Newtonian gravitational potential. 
The Jacobi equation expresses the linear evolution of the deviation
between two neighboring geodesic worldlines with
negligible relative motion. Equation~\eqref{4} has been employed
extensively in the treatment of tidal effects in general 
relativity (e.g. Stewart \& Walker~\cite{new7}; Mashhoon \cite{3,4}; 
Carter \& Luminet~\cite{CL}; Ivanov \& Novikov~\cite{IN}).
Detailed calculations have revealed that the \emph{dominant}
 tidal effects of massive
bodies are rather similar to their Newtonian counterparts 
except possibly \emph{very} close
to  black holes.

Equation \eqref{eq37} follows from the timelike character
 of particle motion; that is, the requirement that 
$ g_{\mu \nu} U^\mu U^\nu = -1$ leads to  \eqref{eq37}
when we use the terms in the expansion of the metric tensor given 
explicitly in  \eqref{1}. 
The modified Lorentz factor is positive and approaches 
infinity if the timelike geodesic approaches a null geodesic. 
It follows from  \eqref{eq37} that $V^2 < 1$ along the reference geodesic; 
elsewhere in the Fermi frame, however, $V^2$ is simply constrained by the 
requirement that the right-hand side of  \eqref{eq37} be positive.

The quasi-inertial Fermi coordinate system has been widely used in the
theory of general relativity; for example, it is 
employed by Ehlers \& Rindler~(\cite{new10}) to study
the local bending of light in a gravitational field.

\section{Tidal acceleration in Kerr spacetime}
In this Letter we point out
the major consequences of \eqref{3} and \eqref{eq37} 
for ultrarelativistic motion in the field of a rotating source represented
by the exterior Kerr spacetime.
The Kerr metric can be expressed in Boyer-Lindquist coordinates
$(t,r,\vartheta,\phi)$ as 
\begin{eqnarray}\label{5}
\nonumber ds^2&=&-dt^2+\Sigma (\frac{1}{\Delta}dr^2+d\vartheta^2)+(r^2+a^2)\sin^2\vartheta\,d\phi^2\\
&&{}+2GM\frac{r}{\Sigma}(dt-a\sin^2\vartheta\,d\phi)^2,
\end{eqnarray}
where $\Sigma=r^2+a^2\cos^2\vartheta$ and $\Delta=r^2-2GMr+a^2$.
Here $M$ is the mass, $J=Ma$ is the angular momentum of the source and we use 
units such that $c=1$. We take $\mathcal{O}$ to be a test particle moving
along the axis of rotation $(\vartheta=0)$ on an escape trajectory
such that it reaches infinity with a speed that is small compared to unity; 
in fact, we set this speed equal to
zero, since our results turn out to be insensitive to it.
The worldline of the reference particle is thus characterized by
\begin{equation}\label{6}
\frac{dt}{d\tau} = \frac{r^2+a^2}{r^2-2GMr+a^2},\quad 
\frac{dr}{d\tau }=\sqrt{\frac{2GMr}{r^2+a^2}}.
\end{equation}
We assume that at $t=\tau=0$, the reference particle starts
from $r=r_0>\sqrt{3}\, a$ along the rotation axis and
eventually escapes to infinity in the absence of other interactions.
Given this reference trajectory, 
it is then necessary to choose
the spatial axes of the tetrad frame $\lambda^\mu_{\;\;(\alpha)}$.
 Taking $\lambda^\mu_{\;\;(3)}$ to be along the direction of motion,
axial symmetry about this direction implies that there is rotational
degeneracy in the choice of $\lambda^\mu_{\;\;(1)}$
and $\lambda^\mu_{\;\;(2)}$. Once a triad is chosen, however,
it is parallel propagated along the reference worldline.
The projection of the Riemann tensor on this tetrad, as in \eqref{2},
is given by
a symmetric $6\times 6$ matrix with 
indices that range over the set $\{01,02,03, 23, 31,12\}$:
\begin{equation}\label{7}
\left[ \begin{array}{cc}
E & B \\
 B & -E
\end{array}\right ],
\end{equation}
where $E$ and $B$ are symmetric and traceless
$3\times 3$ matrices that represent
the electric and magnetic parts of the curvature tensor.
It turns out (Mashhoon, McClune \& Quevedo~\cite{n8})
 that $E$ and $B$ are both diagonal and parallel:
$E/k=B/q=\diag (-\frac{1}{2},-\frac{1}{2},1)$,
where
\begin{equation}\label{8}
k=-2GM\frac{r(r^2-3a^2)}{(r^2+a^2)^3},
\quad q=2\; GM a\frac{3r^2-a^2}{(r^2+a^2)^3}.
\end{equation}
Integrating the equation of motion of the reference
geodesic~\eqref{6}, we find $r(T)$ and its substitution in
equation~\eqref{8} gives the electric and magnetic curvatures
$k(T)<0$ and $q(T)>0$, respectively.
The class of such reference particles provides an ambient medium 
around the source. 
In effect, this medium provides a \emph{local}
dynamic connection with the rest frame of the source.

The generalized Jacobi equation~\eqref{3} in the case
under consideration takes the form
\begin{eqnarray}
\label{9}
\nonumber &&\ddot{X}-\frac{1}{2}kX
\Big[ 1-2\dot{X}^2+\frac{2}{3}
(2 \dot{Y}^2-\dot{Z}^2)\Big] + \frac{1}{3} k\dot{X}( 5Y\dot{Y} -7Z\dot{Z})\\
&&\qquad + q[\dot{X}\dot{Y}\dot{Z} X-\dot{Z}Y(1+\dot{X}^2)-2\dot{Y}Z]=0,\\
\label{10}
\nonumber &&\ddot{Y}- \frac{1}{2} kY
\Big[ 1-2\dot{Y}^2 +\frac{2}{3}(2 \dot{X}^2-\dot{Z}^2)\Big]
 +\frac{1}{3} k\dot{Y} (5X\dot{X} -7Z\dot{Z})\\
&&\qquad-q [\dot{X}\dot{Y}\dot{Z}Y-\dot{Z}X(1+\dot{Y}^2)-2\dot{X}Z]=0,\\
\label{11}
\nonumber &&\ddot{Z} + kZ
\Big[ 1-2\dot{Z}^2+\frac{1}{3} (\dot{X}^2+\dot{Y}^2)\Big] +\frac{2}{3}
k\dot{Z} (X\dot{X} +Y\dot{Y})\\
&&\qquad-q(X\dot{Y}-\dot{X} Y)(1-\dot{Z}^2)=0,
\end{eqnarray}
where $\dot{X}=dX/dT$, etc.
Let us first concentrate on the motion of $\mathcal{P}$ along the rotation axis; that is,
$X(T)=Y(T)=0$ for all $T\ge 0$.
In this case, equations~\eqref{9}--\eqref{11} reduce to
\begin{equation}\label{12}
\frac{d^2Z}{dT^2}+k(1-2\dot Z^2)Z=0.
\end{equation}
The character of the solutions of this equation depends
drastically on whether the relative initial velocity along the $Z$-direction
has magnitude below or above the critical speed $1/\sqrt{2}\,$.
For initially \emph{infrarelativistic} motion (i.e. $\dot Z^2<1/2$),
equation~\eqref{12} implies that the particle accelerates
until it asymptotically approaches the critical speed.
For initially \emph{ultrarelativistic} motion  (i.e. $\dot Z^2>1/2$),
however, the particle strongly decelerates \emph{regardless of its
initial Lorentz factor} $\Gamma_0>\sqrt{2}$ and asymptotically
approaches the critical speed $1/\sqrt{2}\,$.
This deceleration phenomenon is consistent with the observations of
Galactic ``superluminal'' jets (Fender~\cite{F}).

\begin{figure}
\centerline{\psfig{file=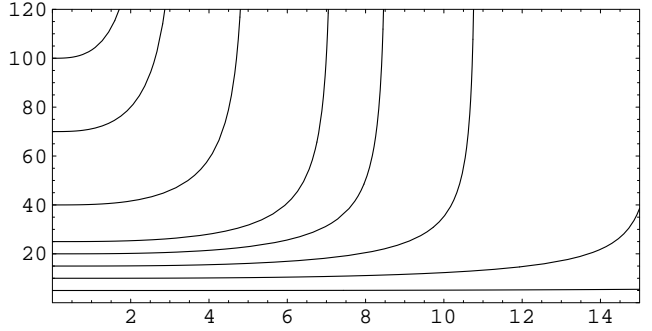, width=20pc}}
\caption{Plot of the Lorentz factor $\Gamma=(1-\dot X^2- kX^2/2)^{-1/2}$ versus $T/(GM)$ based on
integration of equation~\eqref{14} with
initial data $X=0$ at $T=0$ with
$\dot X(0)=(\Gamma_0^2-1)^{1/2}/\Gamma_0$ corresponding respectively
to $\Gamma_0=5$, 10, 15, 20, 25, 40, 70 and 100. In this plot $a/(GM)=1$
and $r_0/(GM)=15$. The graph illustrates acceleration of
the particle such that $\Gamma$ essentially
approaches infinity at $T/(GM)\approx 11$, 9, 7, 5, 3 and 2.6 
for $\Gamma_0=15$, 20, 25, 40, 70 and 100, respectively.  \label{fig:2}}
\end{figure}
Let us next consider motion along the $X$-axis; that is,  $Y(T)=Z(T)=0$
for all $T\ge 0$. 
Then equations~\eqref{9}--\eqref{11} reduce to
\begin{equation}\label{14}
\frac{d^2X}{dT^2}-\frac{1}{2} k(1-2\dot X^2)X=0,
\end{equation}
which turns out to be a generic equation for motion
perpendicular to the symmetry axis as a consequence of the rotational
symmetry of equations~\eqref{9}--\eqref{11} about the $Z$-axis.
For initially infrarelativistic motion, it follows from \eqref{14}
that the particle slowly decelerates, while for initially
ultrarelativistic motion, the particle strongly accelerates as depicted 
in Fig.~\ref{fig:2}. The latter situation is remarkable, as it can lead
to $\Gamma\to\infty$ in the absence of other interactions. 
It is clear from Fig.~\ref{fig:2} 
that in some cases the Fermi coordinate system could break
down before $\Gamma \to \infty$. 
The breakdown in our dynamics for $\Gamma \to \infty$ is logically
distinct from the kinematic breakdown of the Fermi coordinate system
beyond $|{\bf X}|\sim\mathcal{R}$. These tentative conclusions follow 
from our simple
scenario that takes into account only the quadratic terms in equation~\eqref{1}; in
principle, the whole infinite series must be considered.
That is, the geodesic equation of motion of $\mathcal{P}$ involves an infinite
series of tidal terms; the breakdown occurs when instead of the infinite set
of tidal terms only the first-order terms are taken into account. 
Our  analytical and numerical studies of 
the general structure of this infinite series have led us to the working 
hypothesis that for ultrarelativistic particles, finite but significant tidal
acceleration takes place normal to the jet direction
(Chicone \& Mashhoon~\cite{new20}).

The breakdown of our simple theoretical scheme as $\Gamma \to \infty$
implies
that the influence of such a particle on the background spacetime cannot be
neglected. A more complete treatment should take this influence into
account, thereby moderating the singularity that we have encountered.
Moreover, the tidal energy exchange resulting in the
deceleration/acceleration phenomena must be augmented to take due account of
gravitational radiation energy emitted by the particles.  This energy, as well as
the corresponding electromagnetic radiation energy emitted by the electrically
charged particles,  is in the form of long-wavelength radiation of frequency
proportional to $( 2GM )^{-1} = 10^5 ( M_\odot / M )\, \mbox{Hz}$.
The tidal acceleration of charged particles could be strongly affected by the
electromagnetic field configuration around the source.

It is interesting to consider the exchange of energy between the particle
$\mathcal{P}$ and the gravitational field. 
Einstein's principle of equivalence
prevents a local (i.e. pointwise) description of gravitational energy.
From the standpoint of observers at rest with the ambient medium, however, an
initially ultrarelativistic particle would lose energy to the gravitational
field along the jet direction, but would gain tidal energy propagating
outward in the equatorial plane; thus, the gain in gravitational energy by the 
accelerating particles could be essentially balanced with the loss of 
gravitational energy by the particles decelerating along the jets.

For an invariant characterization of these tidal phenomena, consider the
class of fundamental observers $\mathcal{S}$ that are at rest with 
four-velocity
$u^\mu = (W, 0, 0, 0)$ with respect to the Fermi coordinate system. 
Here $W =(-g_{00})^{-1/2}$. Let $\hat\Gamma := - u^\mu U_\mu$ 
be the Lorentz factor of $\mathcal{P}$
measured by the static observers $\mathcal{S}$, 
which are in general accelerated.
It turns out that
\begin{equation}\label{blde}
 \hat\Gamma = (W^{-1} - W g_{0\,i} V^i ) \Gamma.          
\end{equation}
We find that $\hat\Gamma$ behaves essentially as $\Gamma$; e.g., 
in Fig.~\ref{fig:2}, the
plot of $\hat\Gamma$ would be indistinguishable from that of $\Gamma$.

Ultrarelativistic particles are
expected to be produced 
near a highly magnetized rapidly rotating neutron star or 
as a consequence of the complicated
accretion phenomena in the vicinity of an active black hole
(Punsly~\cite{8}; Guthmann et al.~\cite{G}).
For our purposes, we imagine an abundance of such particles
near the poles of the collapsed system. Extensive numerical studies
of equations~\eqref{9}--\eqref{11}
indicate that the deceleration along the rotation
axis extends to a cone with an opening half-angle $\theta$ given
by $\tan\theta\approx\sqrt{2}\,$;
moreover, motion outside the cone corresponds to tidal acceleration
away from the collapsed system. 

Tidal acceleration of ultrarelativistic particles is maximum in the 
equatorial plane and decreases away from it, turning to deceleration at an 
inclination angle of about $35^\circ$. 
Our results regarding tidal acceleration appear to be consistent 
with recent \emph{Chandra} X-ray studies of four neutron stars: 
Crab Pulsar, Vela Pulsar, PSR B1509-58 and SNR G54.1+0.3 
(http://chandra.harvard.edu/photo/chronological.html). 
In each case, there is considerable activity near the equatorial plane of the 
central source. Moreover, a detailed analysis of the \emph{Chandra} X-ray 
images of the Crab Nebula suggests that the equatorial activity is 
somewhat more energetic than the activity along the jets
(Mori et al.~\cite{new15}), 
which is consistent with the tidal acceleration/deceleration of 
ultrarelativistic particles studied in this work. 
The same appears to be the case for the Vela Pulsar
(http://chandra.harvard.edu/photo/2000/vela/index.html).

\section{Discussion}
It is important to emphasize
that the deceleration/acceleration phenomena are generic to
collapsed configurations; the rotation of the central source has a negligible
influence on these results.
In particular, the results presented
in Fig.~\ref{fig:2} are insensitive 
to the assumption that the two particles
initially coincide, or that the central source rotates rapidly. 
Numerical experiments
for initially nearby particles as well as for $a$ in the range $[ 0,GM]$
demonstrate that the
results presented here are robust. 
What matters is that the central source is a neutron star or a
black hole; the largest effects are expected to occur in the latter case.
In either case, tidal deceleration/acceleration can be significant near the source,
where the ultrarelativistic particles are launched; however, a complete
treatment should take due account of the complicated plasma processes 
as well (Punsly~\cite{8}; Guthmann et al.~\cite{G}).

The deceleration/acceleration phenomena occur \emph{relative to the ambient
medium}. That is, the geodesic orbits of the exterior Kerr spacetime under
consideration in this work would not appear to have any extraordinary
features when considered from the standpoint of the static inertial
observers at spatial infinity. The situation is different, however, in the
Fermi system.

The ultrarelativistic particles that are created near a 
gravitationally collapsed system or in the 
accretion process around the system are tidally
decelerated in a cone around the rotation axis of the collapsed system and 
appear as
confined relativistic outflows from the system.
These correspond to jets from some neutron stars 
(http://chandra.harvard.edu/photo/chronological.html) or
X-ray binaries (microquasars) 
in our galaxy that have been extensively 
studied (Guthmann et al.~\cite{G}; Fender~\cite{F}). 
On the other hand, the extremely relativistic particles
that result from tidal acceleration outside the cone may interact
with the ambient medium within the Fermi system thereby transferring their
tidal energy to the ambient particles that may escape from the system
altogether and can appear as extremely energetic cosmic rays. 
Such neutron star or microquasar
particles may account for the inferred 
ultrahigh energy primary cosmic rays with
energies above $10^{20}\, \mbox{eV}$, since the corresponding 
extragalactic particles  would
collide with the cosmic microwave background photons resulting in photopion
production and pair creation (Greisen~\cite{13}; Zatsepin \& Kuzmin~\cite{14}). 
It would therefore be
interesting to search for any correlation between the 
directionality associated
with the highest energy cosmic rays
and the distribution of certain neutron stars and 
microquasars in our galaxy using the Pierre Auger Observatory.

\end{document}